\documentclass[twocolumns]{aa}  

\usepackage{graphicx}

\usepackage{float}
\usepackage{subfig}
\usepackage{txfonts}
\usepackage{url}
\usepackage[switch]{lineno}
\usepackage{datetime}
\usepackage{wrapfig}
\usepackage[usenames,table,dvipsnames]{xcolor}
\usepackage{setspace}
\usepackage[normalem]{ulem}
\usepackage{soul}
\usepackage{hyperref}

\definecolor{blue}{RGB}{66, 153, 233}
\definecolor{red}{RGB}{255, 0, 0}
\definecolor{purple}{RGB}{255, 0, 255}

\begin{document}

   \title{The X-ray synchrotron rims in Cassiopeia A narrow with energy}

   \author{A. Picquenot
          \inst{1}\inst{2}\inst{3}
          \and
          B.~J. Williams
          \inst{2}
          \and
          F. Acero
          \inst{4}
          \and
          B.~T. Guest
          \inst{1}\inst{2}\inst{3}
          }
          
\institute{Department of Astronomy, University of Maryland, College Park, MD 20742 ;
              \and
              X-ray Astrophysics Laboratory NASA/GSFC, Greenbelt, MD 20771
              \and
              Center for Research and Exploration in Space Science and Technology, NASA/GSFC, Greenbelt, MD 20771, USA
              \and
              AIM, CEA, CNRS, Universit\'e Paris-Saclay, Universit\'e de Paris, F-91191 Gif sur Yvette, France
             }

   \date{\today}

 
  \abstract
   {Some young supernova remnants exhibit thin filaments of X-ray synchrotron radiation coinciding with the forward shock due to accelerated electrons interacting with the local magnetic field. The two main models accounting for the radial brightness evolution of these filaments differ in their prediction of the narrowing (or not) of the filaments with increasing photon energy.}
   {In this paper, we report our observation of such a narrowing of the synchrotron filaments in Cassiopeia A at X-ray energies, and how this finding could help in understanding the mechanisms at stake in their formation.}
   {We used a new blind source separation method on the $1$ Ms {\it Chandra} observation of Cassiopeia A, in order to obtain detailed and unpolluted images of the synchrotron emission in three energy bands. We then extracted the profiles of several filaments at the forward shock and the reverse shock to estimate and compare their widths.}
   {We find that there is indeed a narrowing with energy of the synchrotron filaments both at the forward and at the reverse shocks in Cassiopeia A. The energy dependency of this narrowing seems stronger at high energy, which is indicative of a damping effect, confirmed by radio observations.}
   {}


   \maketitle
%

\section{Introduction}

Electrons accelerated in the forward shocks of young supernova remnants (SNR) can emit synchrotron radiation. This emission is mostly seen at radio wavelengths, but can also be seen in X-rays thanks to the fast \citep[$\geq 3000$ km s$^{-1}$, see][]{Aharonian09} shocks of young SNRs. In some cases, the X-ray brightness fades downstream and produces a bright shell-like structure of rims and filaments.

A better understanding of these rims could help constrain the downstream magnetic field structure. Current models use three physical effects to account for the existence of the synchrotron filaments in X-ray: advection (bulk motion of the plasma), diffusion (random motion of electrons at small scales), and magnetic damping (where the X-ray emission reflects the magnetic field morphology). The relative importance of these effects has an influence on the evolution of the filaments widths with energy.

\subsection{The evolution of rim widths with energy}

The models describing the formation of X-ray synchrotron rims can mostly be divided into two main categories, depending on how they take magnetic damping effects into account. In the ``loss-limited'' models, the magnetic field is considered constant over the width of the rim. Electrons only travel a certain distance until they lose enough energy through diffusion and advection for their radiation to drop below the X-ray band \citep[][]{Vink_2003,Bamba_2005,Parizot2006}. If the magnetic field is constant downstream, most energetic electrons will radiate and cool more quickly through advection, which results in a narrowing of the synchrotron rims with energy. However, diffusion will dilute this effect: more energetic electrons may diffuse further than would be expected from pure advection, weakening the energy dependence of the widths at higher energies \citep[][]{Araya_2010,Ressler_2014}. Hence, "loss-limited" models predict a narrowing of the X-ray synchrotron filaments, with a weakening of the energy dependence with energy. 

``Loss-limited'' models require a strong magnetic field amplification to successfully account for the formation of thin rims. \cite{Pohl_2005} proposed that the rim profiles could reflect the magnetic field morphology, adding a magnetic damping mechanism to the effects of diffusion and advection. The magnetic field could be damped downstream of the shock and prevent electrons from radiating efficiently, thus reducing synchrotron flux at all wavelengths. Thus, a damping mechanism is supposed to produce relatively energy-independent rim widths below a threshold energy and may decrease or increase once advection and/or diffusion controls rim widths \citep{Tran_2015}. Hence, the evolution of these widths with energy is a constraint on the magnetic field amplification and the amount of damping. However, there are various physically possible damping mechanisms, so the relationship between the filaments widths and the actual magnetic field variations is not straightforward.

\cite{Ressler_2014} observed synchrotron narrowing in SN$1006$ and concluded that it was too strongly energy-dependent to be well described by the damping mechanism only: the damping lengths would need to be larger than the synchrotron-loss lengths. \cite{Tran_2015} conducted a similar study in Tycho's SNR. They proposed a range of radio and X-ray rim profiles showing the influences of both the magnetic field and the damping length. Their work highlights the similarity between the effects of a strong magnetic field and a small damping length on lowering the dependence in energy of the rim widths, showing that it is difficult to probe a model with no further information on the magnetic field. In the same paper, a moderate narrowing of the rim widths was found in Tycho's SNR in X-rays, but this was insufficient to constrain the model. However, ``loss-limited'' models alone cannot account for the formation of thin radio filaments: hydrodynamic models with diffusive shock acceleration cannot produce radio profiles with narrow rims in a purely advected magnetic field \citep{Cassam_Chenai_2007,2014ApJ...783...33S}. The observation of thin radio filaments hence proved that there is a damping effect in Tycho, which is not necessarily sufficient to account for the narrowing in X-rays.

\subsection{Thin X-ray rims in Cassiopeia A}

Cassiopeia A (hereafter, Cas A) is among the most studied astronomical objects at X-ray wavelengths. It benefits from extensive observations (about $3$ Ms in total with {\it Chandra}) and it is surrounded by a synchrotron shell showing filamentary structures, making it an ideal laboratory to investigate the potential narrowing of synchrotron rims with energy.

\begin{figure}
\centering
\subfloat{\includegraphics[width = 9.8cm]{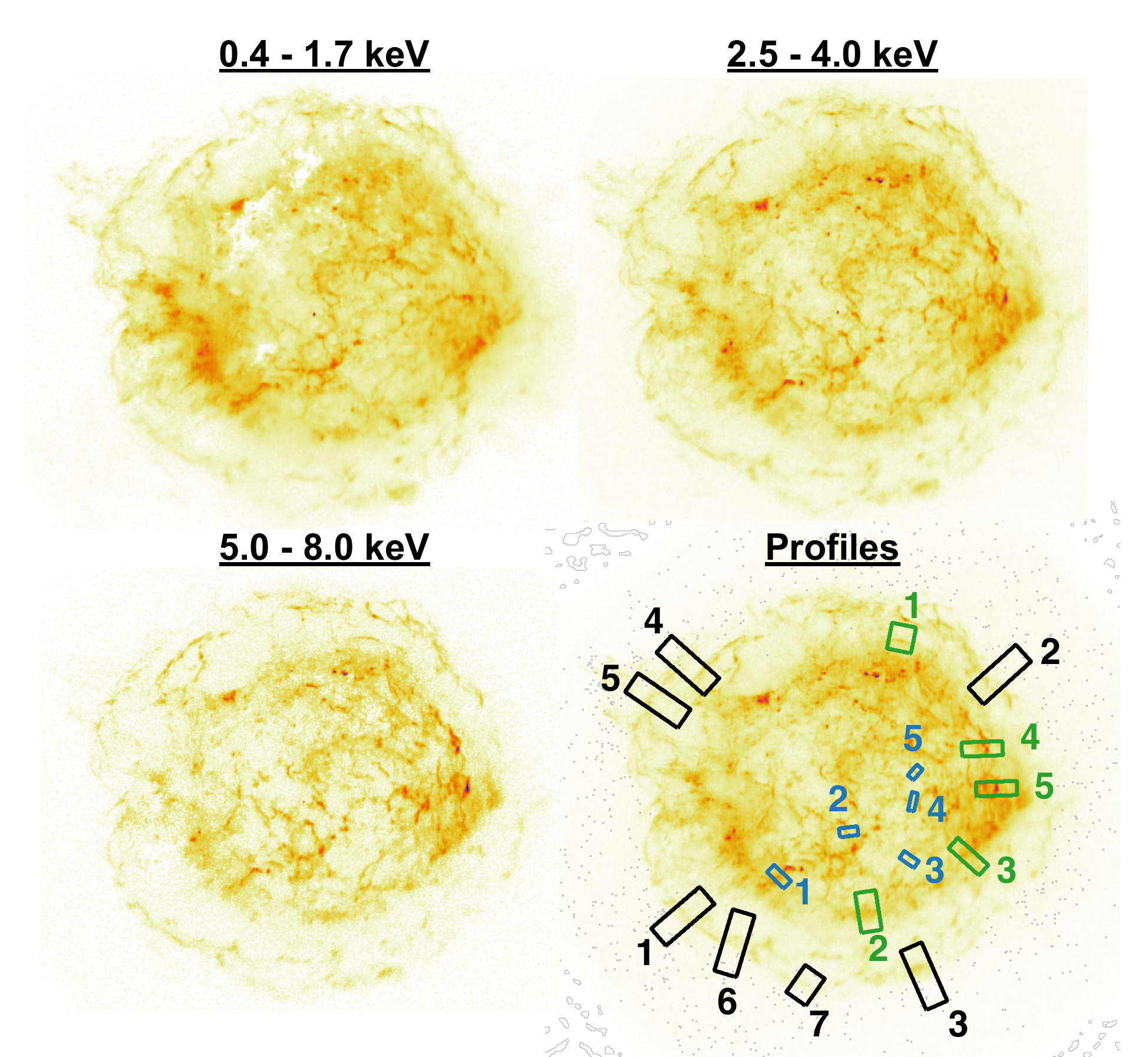}}
\caption{Images of the synchrotron emission obtained with pGMCA in three energy bands, with square root scaling. On the bottom right, the boxes used to define the filament profiles. In black, profiles at the forward shock. In green, profiles at the reverse shock. In blue, unidentified profiles.}
\label{fig:images_sync}
\end{figure}

 \cite{Araya_2010} investigated some properties of the synchrotron rims visible at the forward shock in the $1$ Ms {\it Chandra} observation of Cas~A, including a comparison between
the linear profiles of some filaments in the $0.3$-$2.0$ keV, $3.0$-$6.0$ keV and $6.0-10.0$ keV energy bands. They only found a slight difference in the widths between the $0.3$-$2.0$ keV and the $3.0$-$6.0$ keV images, but none between the $3.0$-$6.0$ keV and the $6.0$-$10.0$ keV images, as shown in Table $1$ of their paper. They also found that the shape of the decline in emission downstream of the shock was similar to that upstream, contrary to model predictions.

Here, we obtained more detailed images by using a new method to retrieve accurate maps of the synchrotron emission around different energy bands. This method is based on the General Morphological Components Analysis \citep[GMCA, see][]{bobin15}, a blind source separation algorithm that was introduced for X-ray observations by \cite{picquenot:hal-02160434}. It can disentangle spectrally and spatially mixed components from an X-ray data cube of the form $(x,y,E)$. The new images thus obtained suffer less contamination by other components, such as thermal emission from lines or continuum. An updated version of this algorithm, the pGMCA \citep[see][]{9215040}, has been developed to take into account the Poissonian nature of X-ray data. It was first used on Cas~A data to probe the three-dimensional morphological asymmetries in the ejecta distribution \citep{Picquenot_2021}, and proved perfectly suited for producing clear, detailed and unpolluted images of both the ejecta and the synchrotron at different energies. Thanks to these new images, we are able to study the profiles of some filamentary structures associated with the forward shock, as well as find some associated with the reverse shock. We labelled ``upstream'' and ``downstream'' the sides of the filament profiles according to their location relative to the shock they are associated with. However, the widths of the profiles do not necessarily correspond to the actual width of the upstream and downstream shock as projection effects might have an effect.

This paper is structured as follows. In Section \ref{sect:define-prof}, we will show the images of the synchrotron we used, and the way the filaments profiles are defined. In Section \ref{sect:gaussian}, we will present a way to quantify the narrowing of the filaments, and discuss our results.

\begin{figure}
\centering
\subfloat{\includegraphics[width = 9cm]{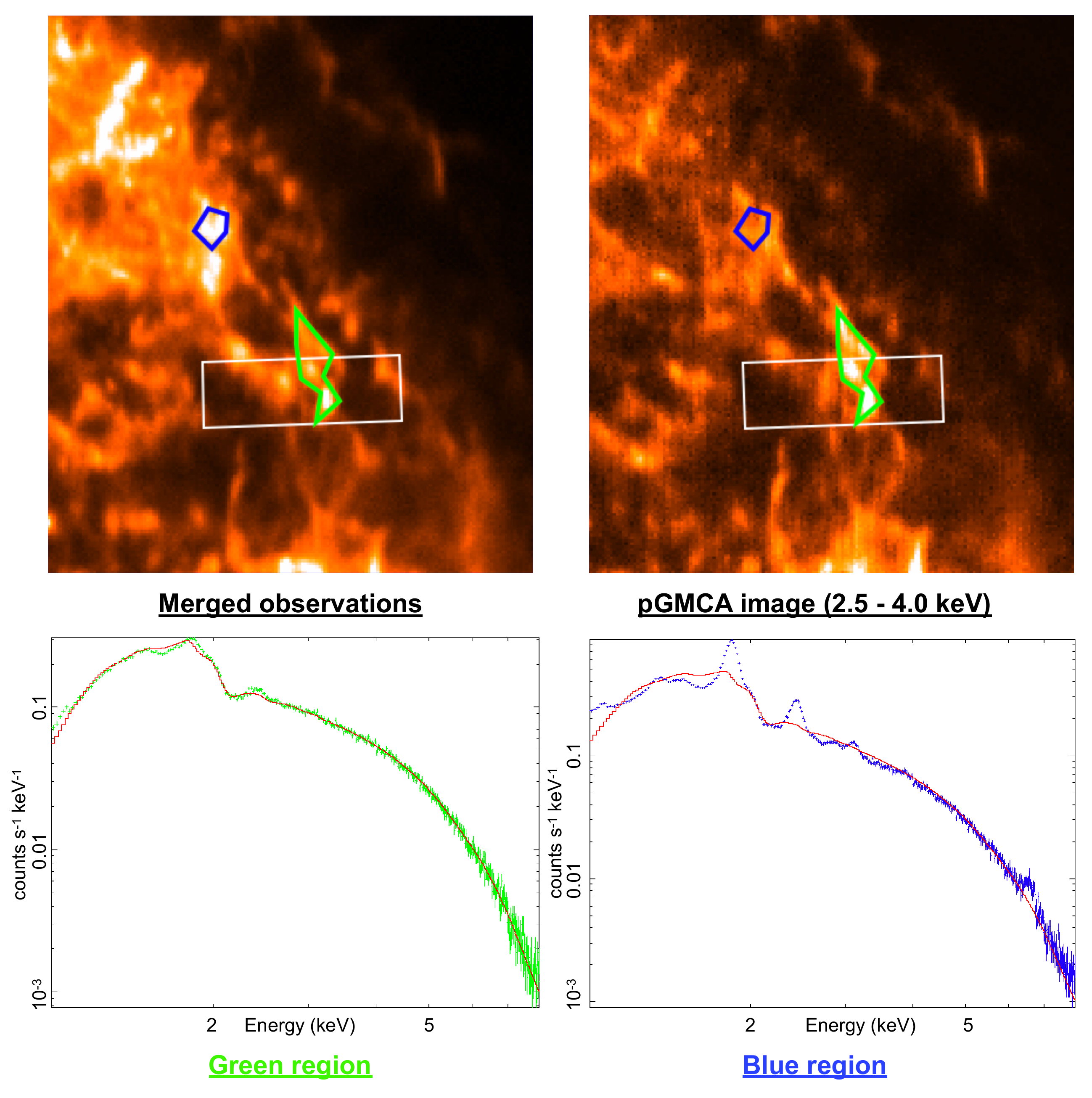}}
\caption{At top left, the northeast region of Cas~A from our merged observations. On the top right, the image of the synchrotron retrieved by our method on the $2.5$ - $4$ keV band. In both, the blue and green contours show the regions of extraction, and the white rectangle shows the 4th reverse shock box from Fig.~\ref{fig:images_sync}. On the bottom, the extracted spectra from both regions, both fitted with a simple \texttt{phabs*powerlaw} model in \textit{Xspec} in red.}
\label{fig:images_filament}
\end{figure}

\section{Using pGMCA to probe the synchrotron rims widths in Cas~A}
\label{sect:define-prof}

\begin{figure*}[ht!]
\centering
\subfloat{\includegraphics[width = 16cm]{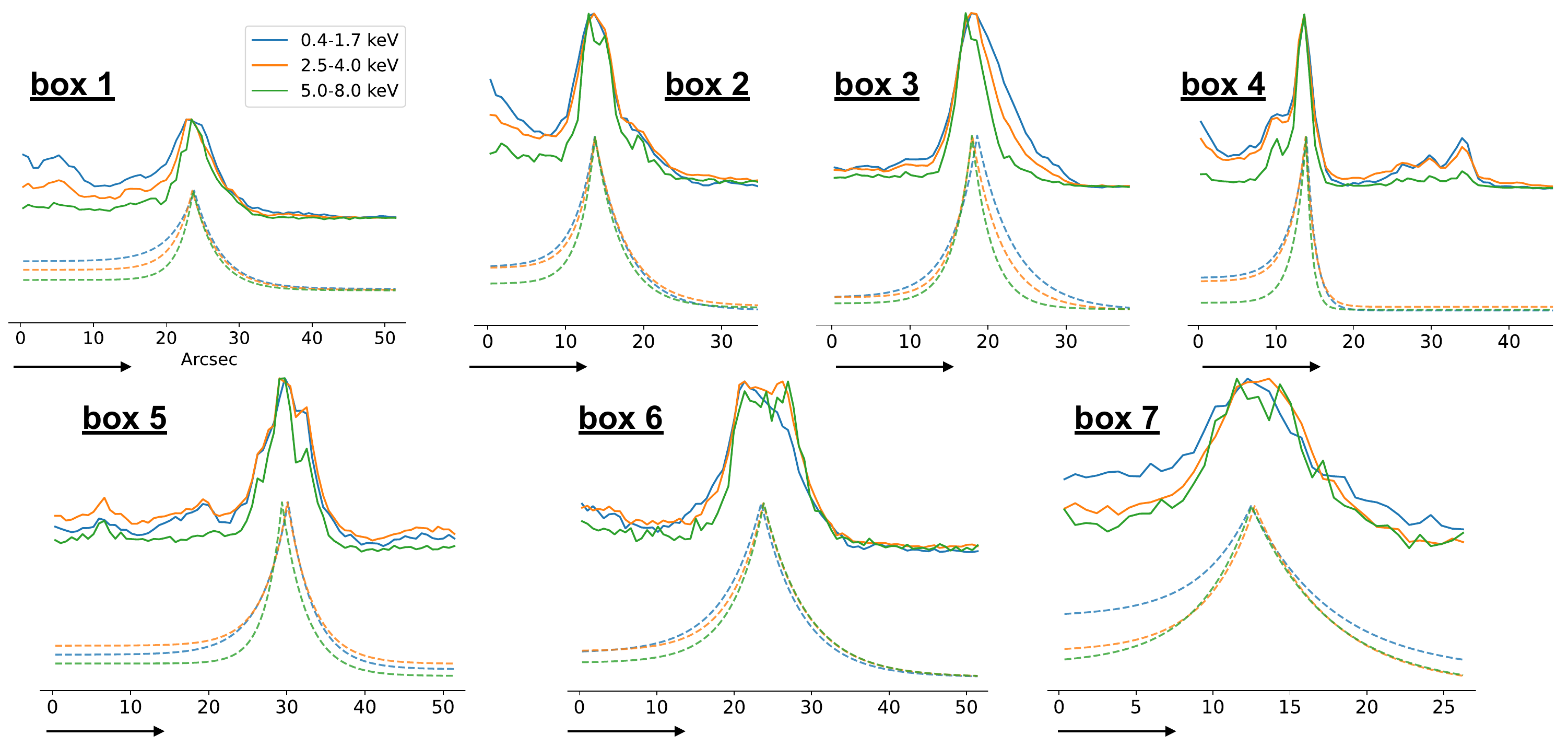}}
\caption{Linear profiles at the forward shock along the black boxes presented in Fig.~\ref{fig:images_sync}, and the models fitted to describe their widths, offset for clarity. The profiles and the models are normalized so that the peaks coincide, and the radius is given in arcsec. The arrows show the direction of the forward shock (upstream is right of the plot).}
\label{fig:profilesFS}
\end{figure*}

\begin{figure*}[ht!]
\centering
\subfloat{\includegraphics[width = 16cm]{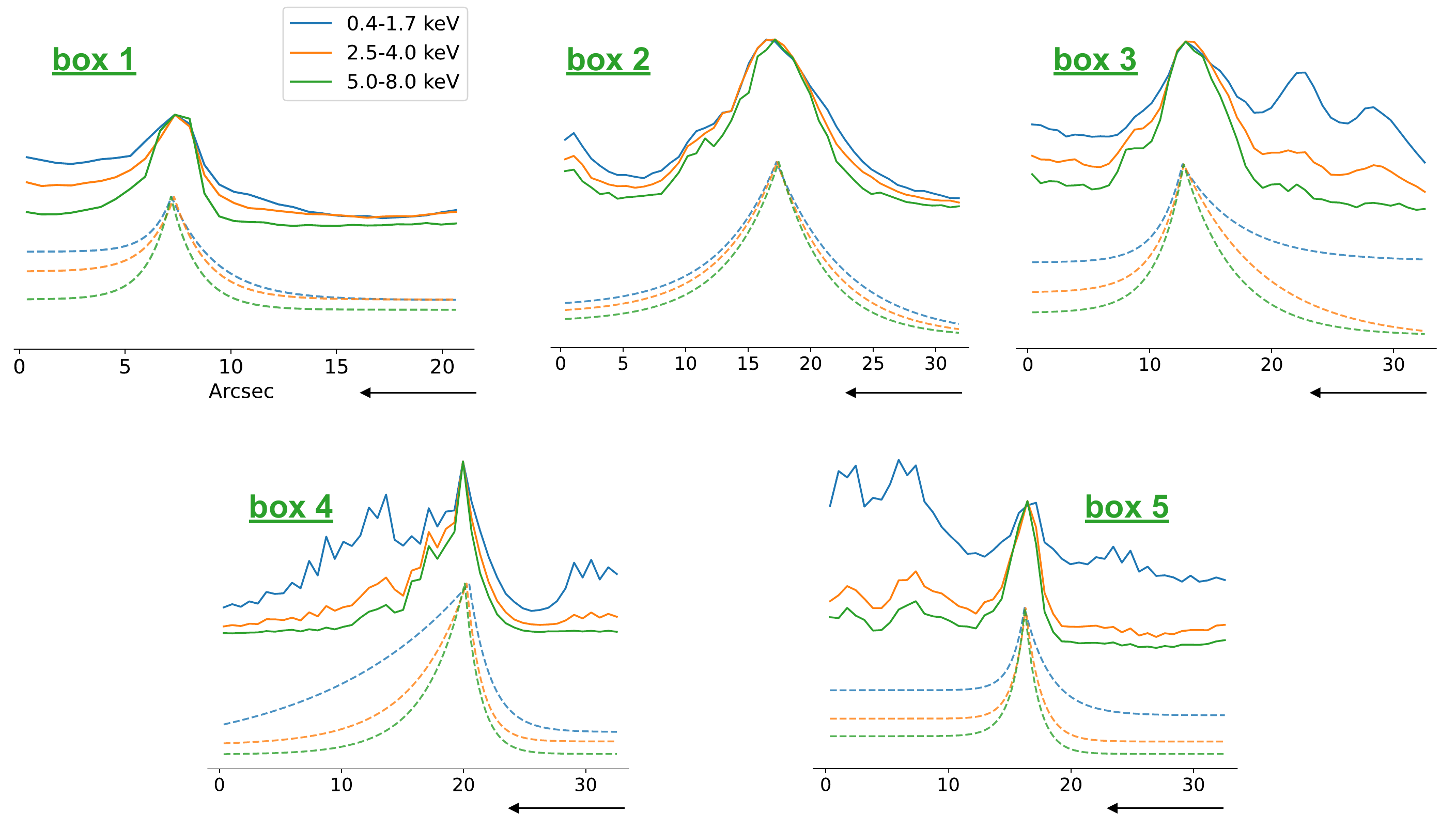}}
\caption{Linear profiles at the reverse shock along the green boxes presented in Fig.~\ref{fig:images_sync}, and the models fitted to describe their widths, offset for clarity. The profiles and the models are normalized so that the peaks coincide, and the radius is given in arcsec. The arrows show the direction of the reverse shock (upstream is left of the plot). }
\label{fig:profilesRS}
\end{figure*}
Being one of the brightest sources in the X-ray sky, Cas~A is the perfect extended source to showcase the capabilities of pGMCA. Cas~A benefits from years of extensive observations, and offers high statistics and strongly overlapping components that the algorithm is well suited to disentangle. Here, we  used it to obtain three images of the synchrotron at different wavelengths, from which we will derive the linear profiles of some wisely chosen synchrotron rims.

\subsection{Images definition}

\begin{figure*}[ht!]
\centering
\subfloat{\includegraphics[width = 16cm]{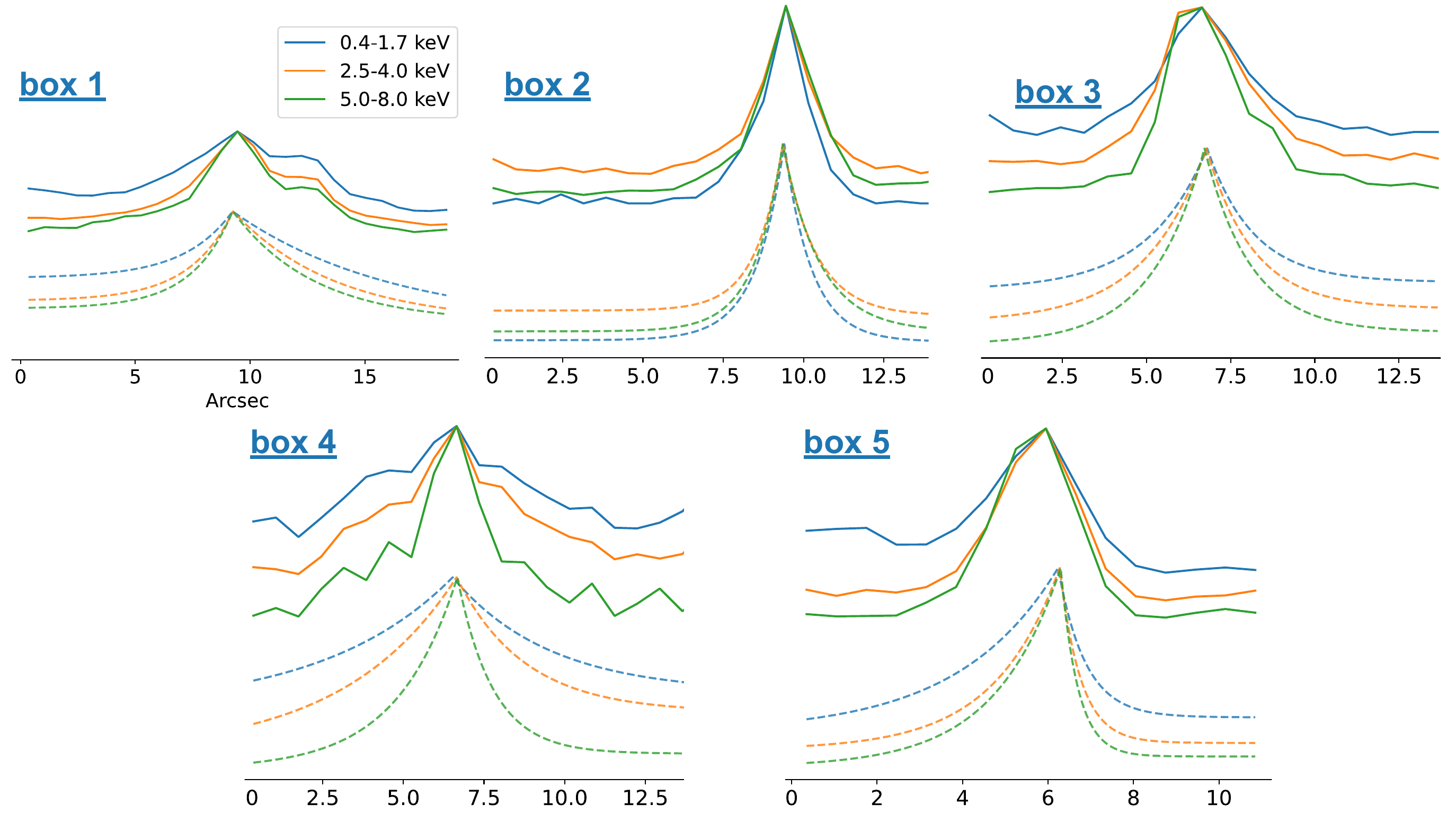}}
\caption{Unidentified linear profiles along the blue boxes presented in Fig.~\ref{fig:images_sync}, and the models fitted to describe their widths, offset for clarity. The profiles and the models are normalized so that the peaks coincide, and the radius is given in arcsec.}
\label{fig:profilesother}
\end{figure*}
For our study, we used {\it Chandra} observations of the Cas~A SNR, which was observed with the ACIS-S instrument in 2004 for a total of 980 ks \citep[][ ObsId 4634, 4635, 4636, 4637, 4638, 4639, 5196, 5319 and 5320]{hwang04}. We used only the 2004 data set to avoid the need to correct for proper motion across epochs. The event lists from all observations were merged in a single data cube. The spatial bin size is the native {\it Chandra} bin size of $0.5$ arcseconds, in order to produce images as detailed as possible. As we are not interested in the spectral lines, we chose a spectral bin size of $58.4$ eV to keep a good number of counts in every pixels.

We applied the pGMCA algorithm on three bandwidths: between $0.4$ and $1.7$ keV, between $2.5$ and $4$ keV and between $5$ and $8$ keV. We chose these bandwidths on different criteria: large enough for pGMCA to work properly, not too large in order not to affect a possible dependency in energy of the filament widths, and we tried as much as possible to avoid any line emission in the same bandwidths. This last criterion could not be fulfilled at lower energies, and some pollution from other emission can be seen in the first image displayed in Fig.~\ref{fig:images_sync}, particularly in the southwest. The two other images seem clear and unpolluted, even with a square root  scaling. They probably constitute the most detailed and accurate maps of the synchrotron in X-rays in Cas A to this day, especially at low energy.

These images present filamentary structures throughout the ejecta, including some following the known layout of the reverse shock. In order to assess the non-thermal nature of these filaments, we extracted the spectrum of one and fitted a \texttt{phabs*powerlaw} model in \textit{Xspec}. The results are shown in Fig.~\ref{fig:images_filament}, together with a spectrum extracted from a bright region with low synchrotron level fitted with the same model. The filament presents a spectrum that can be well-fit with a simple power-law nonthermal model, while the other region obviously cannot. This is consistent with the results from \cite{Helder08}, where synchrotron emission from filaments inside of the remnant was attributed to the reverse shock.

\begin{table*}
    \centering
    
    Forward shock (Fig.~\ref{fig:profilesFS}) \\
    \renewcommand{\arraystretch}{1.4}
    \begin{tabular}{c c c c c c}
      \hline
      \hline
     & & & FWHM (arcsec) & & \\ \cline{3-5}
    Box & Image & Downstream & Upstream & Mean & $m_E$ \\ 
    \hline 
 & $0.4$-$1.7$ keV & $4.46 \pm 0.42$ & $5.02 \pm 0.26$ & $4.74 \pm 0.25$ & \\ 
 1 & $2.5$-$4.0$ keV & $3.39 \pm 0.23$ & $5.06 \pm 0.26$ & $4.22 \pm 0.17$ & $-0.06 \pm 0.04$ \\ 
 & $5.0$-$8.0$ keV & $2.72 \pm 0.14$ & $4.42 \pm 0.19$ & $3.57 \pm 0.12$ & $-0.24 \pm 0.08$ \\ 
\hline 
 & $0.4$-$1.7$ keV & $3.52 \pm 0.41$ & $5.62 \pm 0.30$ & $4.57 \pm 0.25$ & \\ 
 2 & $2.5$-$4.0$ keV & $3.04 \pm 0.29$ & $5.55 \pm 0.28$ & $4.30 \pm 0.20$ & $-0.03 \pm 0.04$ \\ 
 & $5.0$-$8.0$ keV & $2.76 \pm 0.35$ & $4.71 \pm 0.41$ & $3.73 \pm 0.27$ & $-0.20 \pm 0.13$ \\ 
\hline 
 & $0.4$-$1.7$ keV & $4.52 \pm 0.31$ & $6.50 \pm 0.29$ & $5.51 \pm 0.21$ & \\ 
 3 & $2.5$-$4.0$ keV & $3.33 \pm 0.16$ & $5.36 \pm 0.21$ & $4.35 \pm 0.13$ & $-0.13 \pm 0.03$ \\ 
 & $5.0$-$8.0$ keV & $3.22 \pm 0.26$ & $3.52 \pm 0.29$ & $3.37 \pm 0.20$ & $-0.37 \pm 0.09$ \\ 
\hline 
 & $0.4$-$1.7$ keV & $3.08 \pm 0.25$ & $1.61 \pm 0.12$ & $2.34 \pm 0.14$ & \\ 
 4 & $2.5$-$4.0$ keV & $2.82 \pm 0.25$ & $1.63 \pm 0.17$ & $2.23 \pm 0.15$ & $-0.03 \pm 0.05$ \\ 
 & $5.0$-$8.0$ keV & $2.25 \pm 0.15$ & $1.04 \pm 0.13$ & $1.64 \pm 0.10$ & $-0.44 \pm 0.13$ \\ 
\hline 
 & $0.4$-$1.7$ keV & $4.83 \pm 0.42$ & $4.46 \pm 0.39$ & $4.65 \pm 0.29$ & \\ 
 5 & $2.5$-$4.0$ keV & $4.20 \pm 0.42$ & $4.59 \pm 0.45$ & $4.39 \pm 0.31$ & $-0.03 \pm 0.05$ \\ 
 & $5.0$-$8.0$ keV & $2.92 \pm 0.22$ & $4.28 \pm 0.29$ & $3.60 \pm 0.18$ & $-0.29 \pm 0.12$ \\ 
\hline 
 & $0.4$-$1.7$ keV & $6.80 \pm 0.80$ & $7.22 \pm 0.51$ & $7.01 \pm 0.47$ & \\ 
 6 & $2.5$-$4.0$ keV & $5.87 \pm 0.82$ & $7.72 \pm 0.57$ & $6.80 \pm 0.50$ & $-0.02 \pm 0.05$ \\ 
 & $5.0$-$8.0$ keV & $5.66 \pm 0.89$ & $7.74 \pm 0.68$ & $6.70 \pm 0.56$ & $-0.02 \pm 0.16$ \\ 
\hline 
 & $0.4$-$1.7$ keV & $4.54 \pm 0.59$ & $7.42 \pm 0.83$ & $5.98 \pm 0.51$ & \\ 
 7 & $2.5$-$4.0$ keV & $4.25 \pm 0.61$ & $6.91 \pm 0.34$ & $5.58 \pm 0.35$ & $-0.04 \pm 0.06$ \\ 
 & $5.0$-$8.0$ keV & $4.81 \pm 1.06$ & $7.18 \pm 0.65$ & $6.00 \pm 0.62$ & $0.10 \pm 0.18$ \\ 
\hline 
 & $0.4$-$1.7$ keV & $4.54 \pm 0.18$ & $5.41 \pm 0.17$ & $4.97 \pm 0.12$ & \\ 
Mean & $2.5$-$4.0$ keV & $3.84 \pm 0.17$ & $5.26 \pm 0.13$ & $4.55 \pm 0.11$ & $-0.05 \pm 0.02$\\ 
 & $5.0$-$8.0$ keV & $3.48 \pm 0.21$ & $4.70 \pm 0.16$ & $4.09 \pm 0.13$ & $-0.21 \pm 0.05$\\ 
  \hline
\end{tabular}
\caption{\label{tab:profilesFS}FWHMs and $m_E$ values derived from the models fitted on the forward shock filaments profiles shown in Fig.~\ref{fig:profilesFS}. $m_E$ values are calculated between the energy band of the line on which they are written and the energy band of the line above. A negative $m_{E}$ indicates a filament width narrowing with increasing energy. The values' means are arithmetic, and the errors are mean square.}
\end{table*}

\subsection{The filaments linear profiles}
\label{sect:profilesFS}

Thanks to the highly detailed images of the synchrotron emission we were able to find using pGMCA, we could investigate the narrowing with energy of the filaments both at the forward shock and, for the first time, at the reverse shock.

In order to compare the widths of the rims, we defined boxes surrounding small regions crossing a filament at the forward shock, and did likewise for the reverse shock. The boxes are shown in Fig.~\ref{fig:images_sync}, and the normalized linear profiles obtained perpendicularly to the rims are presented in Fig.~\ref{fig:profilesFS} for the forward shock, and in Fig.~\ref{fig:profilesRS} for the reverse shock.

We also looked at some filaments whose positions could not allow us to label clearly. A line-of-sight effect is likely at stake, and these filaments could either be attached to the forward or to the reverse shock. The boxes are also shown in Fig.~\ref{fig:images_sync}, in blue, and the linear profiles obtained perpendicularly to the rim are presented in Fig.~\ref{fig:profilesother}.

\section{Quantifying the narrowing of the synchrotron rims}
\label{sect:gaussian}

\subsection{Modeling the filaments linear profiles}

In order to measure the widths of the filament profiles we extracted, we fitted them with a piecewise two-exponential model, as in \cite{Tran_2015}~:

\begin{equation}
 h(r)=  \left\{ \begin{array}{ll} A_d \exp\Big(\frac{r-r_0}{w_d}\Big) + C_d, \quad r < r_0 \\ A_u \exp\Big(\frac{r-r_0}{w_u}\Big) + C_u, \quad r \geq r_0 \end{array} \right. 
\end{equation}

All parameters are free, except for $A_d$ that is fixed to ensure continuity at $r=r_0$~, with $A_d=A_u+(C_u-C_d)$. 

This model is well adapted to describe the sharp peak displayed by most profiles, but some filaments (such as the $6$th or $7$th of the forward shock) present plateaus or gaussian-like features that are not accounted for in this model. However, this model remains overall a good way to estimate the widths of the profiles we extracted without neglecting the asymmetry around the peak, between the upstream and downstream media. The resulting fitted models are displayed in Fig.~\ref{fig:profilesFS} (forward shock), Fig.~\ref{fig:profilesRS} (reverse shock) and Fig.~\ref{fig:profilesother} (other profiles). In some cases, we had to remove secondary peaks in the profiles to focus on the main one when fitting our model.

From the parameters of our model we can derive two Full Widths at Half Maximum (FWHM) for each profile, FWHM$_u=2\ln({2})w_u$ and FWHM$_d=2\ln({2})w_d$, describing the "sharpness" of the profile on both sides, a larger FWHM meaning a wider profile. The mean between these FWHMs gives an estimation of the actual width of the profile. We also define $m_E$~:

\begin{equation}
 m_E=\frac{\ln\big({\text{FWHM}_2/\text{FWHM}_1}\big)}{\ln\big({E_2/E_1}\big)}
\end{equation}

Where FWHM$_1$ and FWHM$_2$ are the mean FWHMs of a same filament in two energy bands, and $E_1$ and $E_2$ are the lower energy values for each energy band. For each filament, we calculate the $m_E$ between the $0.4$-$1.7$ keV and $2.5$-$4.0$ keV energy bands ($E_1=0.4$, $E_2=2.5$), and between the $2.5$-$4.0$ keV and $5.0$-$8.0$ keV energy bands ($E_1=2.5$, $E_2=5.0$). This parameter aims to evaluate the narrowing of the filaments widths and quantify its dependence on energy: positive values mean widening, negative means narrowing, larger values mean higher energy-dependence while weaker values mean weaker energy-dependence.

The FWHM$_u$, FWHM$_d$, mean FWHM and $m_E$ values derived from our fitted models are shown in Table~\ref{tab:profilesFS} (forward shock) Table~\ref{tab:profilesRS} (reverse shock) and Table~\ref{tab:profilesother} (other profiles). As the ``upstream'' and ``downstream'' labels do not make sense for the profiles that were not clearly identified, we named both sides ``right'' and ``left,'' the orientations corresponding to the plots of Fig.~\ref{fig:profilesother}. As there is no straightforward way to estimate the pGMCA algorithm errors, the errors shown in our Tables are the model fitting errors and their propagation in the calculation of the mean FWHM and $m_E$ values.

\begin{table*}
    \centering
    
    Reverse shock (Fig.~\ref{fig:profilesRS}) \\
    \renewcommand{\arraystretch}{1.4}
    \begin{tabular}{c c c c c c}
      \hline
      \hline
     & & & FWHM (arcsec) & & \\ \cline{3-5}
    Box & Image & Downstream & Upstream & Mean & $m_E$ \\        \hline     
 & $0.4$-$1.7$ keV & $2.85 \pm 0.22$ & $1.33 \pm 0.42$ & $2.09 \pm 0.24$ & \\ 
 1 & $2.5$-$4.0$ keV & $1.67 \pm 0.44$ & $2.14 \pm 0.19$ & $1.90 \pm 0.24$ & $-0.05 \pm 0.09$ \\ 
 & $5.0$-$8.0$ keV & $1.49 \pm 0.40$ & $1.79 \pm 0.25$ & $1.64 \pm 0.24$ & $-0.22 \pm 0.28$ \\ 
\hline 
 & $0.4$-$1.7$ keV & $7.74 \pm 0.43$ & $6.55 \pm 0.86$ & $7.15 \pm 0.48$ & \\ 
 2 & $2.5$-$4.0$ keV & $6.28 \pm 0.76$ & $6.43 \pm 0.40$ & $6.36 \pm 0.43$ & $-0.06 \pm 0.05$ \\ 
 & $5.0$-$8.0$ keV & $5.82 \pm 0.65$ & $5.26 \pm 0.34$ & $5.54 \pm 0.37$ & $-0.20 \pm 0.14$ \\ 
\hline 
 & $0.4$-$1.7$ keV & $6.46 \pm 1.42$ & $2.66 \pm 0.37$ & $4.56 \pm 0.73$ & \\ 
 3 & $2.5$-$4.0$ keV & $3.05 \pm 0.43$ & $9.06 \pm 0.71$ & $6.06 \pm 0.42$ & $0.15 \pm 0.10$ \\ 
 & $5.0$-$8.0$ keV & $3.01 \pm 0.41$ & $5.85 \pm 0.48$ & $4.43 \pm 0.32$ & $-0.45 \pm 0.14$ \\ 
\hline 
 & $0.4$-$1.7$ keV & $2.41 \pm 0.51$ & $16.87 \pm 1.11$ & $9.64 \pm 0.61$ & \\ 
 4 & $2.5$-$4.0$ keV & $6.14 \pm 0.61$ & $1.73 \pm 0.27$ & $3.94 \pm 0.33$ & $-0.49 \pm 0.06$ \\ 
 & $5.0$-$8.0$ keV & $4.53 \pm 0.33$ & $1.72 \pm 0.23$ & $3.13 \pm 0.20$ & $-0.33 \pm 0.15$ \\ 
\hline 
 & $0.4$-$1.7$ keV & $2.97 \pm 1.20$ & $1.43 \pm 0.75$ & $2.20 \pm 0.71$ & \\ 
 5 & $2.5$-$4.0$ keV & $1.37 \pm 0.27$ & $1.74 \pm 0.17$ & $1.55 \pm 0.16$ & $-0.19 \pm 0.18$ \\ 
 & $5.0$-$8.0$ keV & $1.49 \pm 0.19$ & $1.52 \pm 0.12$ & $1.50 \pm 0.11$ & $-0.05 \pm 0.18$ \\ 
\hline 
 & $0.4$-$1.7$ keV & $4.49 \pm 0.40$ & $5.77 \pm 0.34$ & $5.13 \pm 0.26$ & \\ 
Mean & $2.5$-$4.0$ keV & $3.70 \pm 0.24$ & $4.22 \pm 0.18$ & $3.96 \pm 0.15$ & $-0.13 \pm 0.05$\\ 
 & $5.0$-$8.0$ keV & $3.27 \pm 0.19$ & $3.23 \pm 0.14$ & $3.25 \pm 0.12$ & $-0.25 \pm 0.08$\\ 
  \hline
\end{tabular}
\caption{\label{tab:profilesRS}FWHMs and $m_E$ values derived from the models fitted on the forward shock filaments profiles shown in Fig.~\ref{fig:profilesRS}. $m_E$ values are calculated between the energy band of the line on which they are written and the energy band of the line above. The values' means are arithmetic, and the errors are mean square.}
\end{table*}

\begin{table*}
    \centering
    
    Other profiles (Fig.~\ref{fig:profilesother}) \\
    \renewcommand{\arraystretch}{1.4}
    \begin{tabular}{c c c c c c}
      \hline
      \hline
     & & & FWHM (arcsec) & & \\ \cline{3-5}
    Box & Image & Left & Right & Mean & $m_E$ \\ 
    \hline 
 & $0.4$-$1.7$ keV & $2.92 \pm 0.70$ & $9.70 \pm 0.59$ & $6.31 \pm 0.46$ & \\ 
 1 & $2.5$-$4.0$ keV & $2.67 \pm 0.38$ & $6.20 \pm 0.96$ & $4.43 \pm 0.51$ & $-0.19 \pm 0.07$ \\ 
 & $5.0$-$8.0$ keV & $2.47 \pm 0.35$ & $4.98 \pm 0.69$ & $3.73 \pm 0.39$ & $-0.25 \pm 0.22$ \\ 
\hline 
 & $0.4$-$1.7$ keV & $1.23 \pm 0.16$ & $1.18 \pm 0.10$ & $1.21 \pm 0.09$ & \\ 
 2 & $2.5$-$4.0$ keV & $1.15 \pm 0.15$ & $1.41 \pm 0.11$ & $1.28 \pm 0.09$ & $0.03 \pm 0.06$ \\ 
 & $5.0$-$8.0$ keV & $1.21 \pm 0.19$ & $1.59 \pm 0.17$ & $1.40 \pm 0.13$ & $0.13 \pm 0.17$ \\ 
\hline 
 & $0.4$-$1.7$ keV & $2.43 \pm 0.31$ & $1.89 \pm 0.28$ & $2.16 \pm 0.21$ & \\ 
 3 & $2.5$-$4.0$ keV & $2.51 \pm 0.51$ & $1.92 \pm 0.52$ & $2.22 \pm 0.36$ & $0.01 \pm 0.10$ \\ 
 & $5.0$-$8.0$ keV & $2.15 \pm 0.46$ & $1.89 \pm 0.62$ & $2.02 \pm 0.38$ & $-0.13 \pm 0.36$ \\ 
\hline 
 & $0.4$-$1.7$ keV & $5.58 \pm 3.53$ & $4.16 \pm 1.59$ & $4.87 \pm 1.93$ & \\ 
 4 & $2.5$-$4.0$ keV & $4.90 \pm 1.47$ & $2.92 \pm 0.82$ & $3.91 \pm 0.84$ & $-0.12 \pm 0.25$ \\ 
 & $5.0$-$8.0$ keV & $2.61 \pm 0.57$ & $1.53 \pm 0.53$ & $2.07 \pm 0.39$ & $-0.92 \pm 0.41$ \\ 
\hline 
 & $0.4$-$1.7$ keV & $3.23 \pm 1.03$ & $0.88 \pm 0.20$ & $2.06 \pm 0.52$ & \\ 
 5 & $2.5$-$4.0$ keV & $2.13 \pm 0.38$ & $0.68 \pm 0.24$ & $1.40 \pm 0.23$ & $-0.21 \pm 0.16$ \\ 
 & $5.0$-$8.0$ keV & $2.15 \pm 0.27$ & $0.56 \pm 0.31$ & $1.35 \pm 0.20$ & $-0.06 \pm 0.32$ \\ 
\hline 
 & $0.4$-$1.7$ keV & $3.08 \pm 0.75$ & $3.56 \pm 0.35$ & $3.32 \pm 0.41$ & \\ 
Mean & $2.5$-$4.0$ keV & $2.67 \pm 0.33$ & $2.63 \pm 0.28$ & $2.65 \pm 0.22$ & $-0.09 \pm 0.07$\\ 
 & $5.0$-$8.0$ keV & $2.12 \pm 0.17$ & $2.11 \pm 0.22$ & $2.11 \pm 0.14$ & $-0.25 \pm 0.14$\\ 
  \hline
\end{tabular}
\caption{\label{tab:profilesother}FWHMs and $m_E$ values derived from the models fitted on the forward shock filaments profiles shown in Fig.~\ref{fig:profilesother}. $m_E$ values are calculated between the energy band of the line on which they are written and the energy band of the line above. The values' means are arithmetic, and the errors are mean square.}
\end{table*}

\subsection{Discussion}

A quick look at the mean FWHMs or at the $m_E$ signs shows that there is indeed a narrowing of the filaments profiles with energy. In all of the seven forward shock profiles, five reverse shock profiles, and five unidentified profiles, only two present a widening between $0.4$-$1.7$ keV and $5.0$-$8.0$ keV: FS $7$ and unidentified profile $2$. In both cases, this widening is so low that the fitting errors allow for the possibility of a narrowing as well. Hence, we can reasonably conclude that the narrowing of the filaments with energy in Cas~A is a global effect, that can be observed on several filaments both at the forward and at the reverse shock. In Sect. \ref{sect:PSF} we will see that this narrowing is not due to the evolution of Chandra's point spread function (PSF) with energy. Future studies could also take into account the possible effects of dust scattering on the observed widths of the filaments. However, in first approximation, scattering effects have an energy dependence in $E^{-2}$ \citep[for example]{https://doi.org/10.48550/arxiv.2209.05261,10.1093/mnras/stv1704}, which is not consistent with our observations. Hence, the narrowing we observe is likely not primarily due to dust scattering.

We can see in the results a common trend regarding the evolution of $m_E$ values: the mean $m_E$ between $2.5$-$4.0$ keV and $5.0$-$8.0$ keV are significantly larger, in absolute, than the mean $m_E$ between $0.4$-$1.7$ keV and $2.5$-$4.0$ keV. While this result is to handle cautiously, given the importance of the errors, the low statistics on which the means are calculated and the non-negligible number of outliers, it would mean that the observed narrowing of the synchrotron rims has a stronger energy-dependence at higher energies than at lower energies. Following \cite{Tran_2015}, this result would be characteristic of a damping mechanism: the rim widths are relatively energy-independent below a threshold energy and increase once advection controls rim widths. This is in apparent contradiction with \cite{Araya_2010}, where a moderate narrowing was observed between $0.3$-$2.0$ keV and $3.0$-$6.0$ keV, but not between $3.0$-$6.0$ keV and $6.0$-$10.0$ keV. However, the energy bands we compared are not the same, and our ``high energy $m_E$'' is defined between $2.5$ and $5.0$ keV, while theirs is between $3.0$ and $6.0$ keV. The highly detailed synchrotron images we obtained with pGMCA may also have an influence, allowing for a more precise profile definition and width measurement with less thermal emission contamination. 

In \cite{Araya_2010}, the model used to describe the forward shock filament profiles predicts a sharp decline upstream, after the peak, that they did not observe. We can see from the FWHM$_u$ and FWHM$_d$ values from Table~\ref{tab:profilesFS} that we did not observe it either. The FWHM$_u$ values even tend to be larger than FWHM$_d$ values, meaning that the profiles are sharper downstream than upstream of the forward shock. On the contrary, we can see in Table~\ref{tab:profilesRS} that at the reverse shock, FWHM$_u$ values are mainly smaller than FWHM$_d$ values, meaning that the profiles are sharper upstream (even though it is not apparent in the mean FWHM$_u$ and FWHM$_d$ values, that are driven by a few outliers). However, we will see in Sect.~\ref{sect:PSF} that this could be linked to the PSF.

Not much can be said to identify our ``other'' profiles as belonging to the forward or to the reverse shock. Although it could be tempting to attempt an identification based on their FWHM on each side, line of sight effects are likely involved and for a filament facing the observer, ``left'' and ``right'' of the peak are not trivially equivalent to ``upstream'' or ``downstream''. Nonetheless, Table~\ref{tab:profilesFS} mostly follows the same trend of narrowing, with stronger energy-dependence at high energies, that we observed on both forward and reverse shock rim profiles.
   

\subsection{Influence of the PSF}
\label{sect:PSF}

Previous studies by \cite{Araya_2010}, \cite{Ressler_2014} and \cite{Tran_2015} did not take into account the PSF in their filaments widths measurements. Yet, the PSF evolves with the energy, and could have an impact on our widths comparison between energy ranges.

\begin{figure}
\centering
\subfloat{\includegraphics[width =9cm]{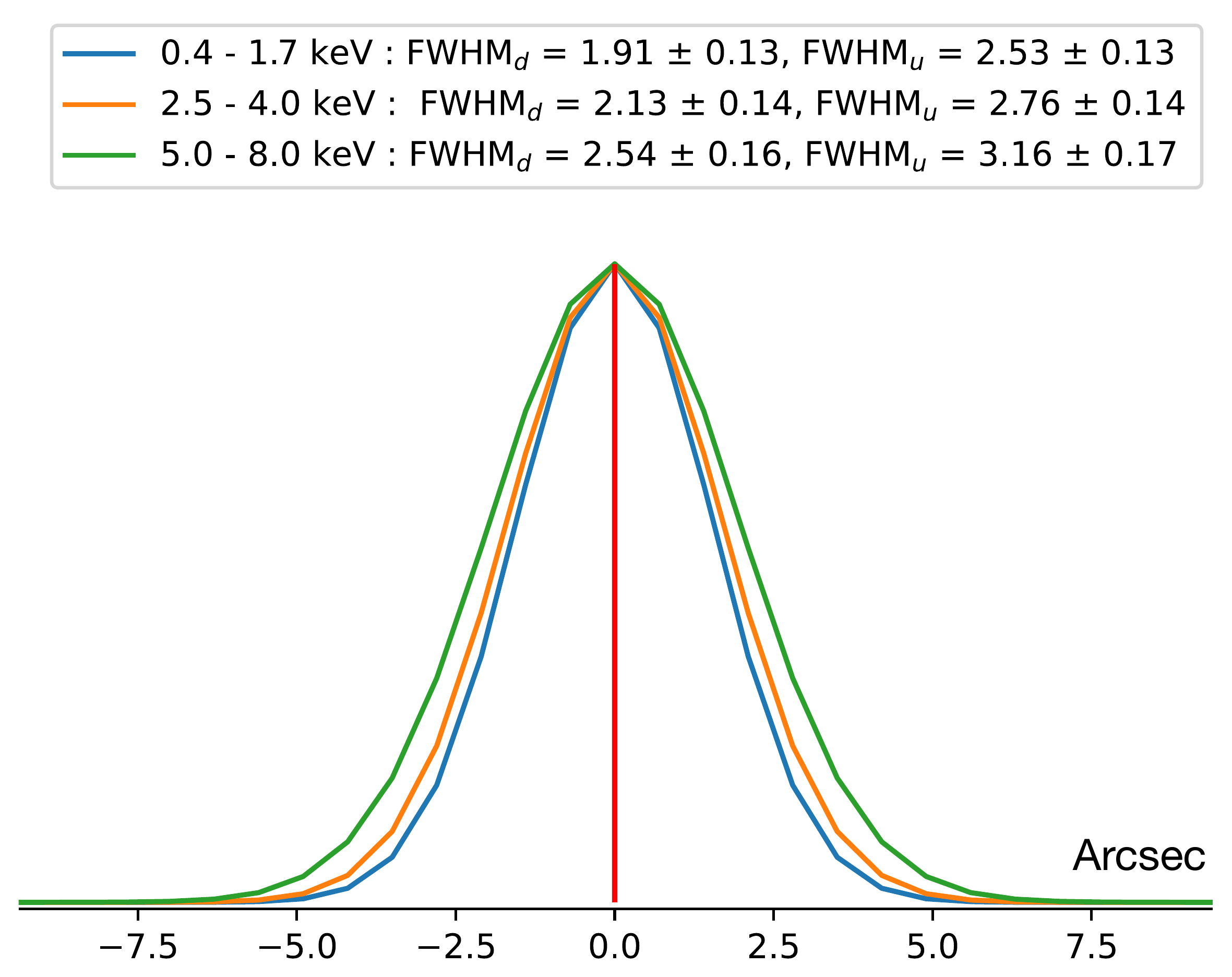}}
\caption{A Dirac delta function is convolved with the PSF profiles along the third FS box for our three energy ranges and normalized. This box was chosen because it was the one along which the PSF varied the most, both spatially and with energy. The FWHM values downstream and upstream are obtained by fitting our piecewise two-exponential model; values are given in arcsec.}
\label{fig:PSF_test}
\end{figure}

In a first attempt to take it into account, we generated the 1-sigma PSF maps of the merged 2004 observations of Cas~A using the \texttt{merge\_obs} routine from CIAO around the $0.4$-$1.7$ keV, $2.5$-$4.0$ keV and $5.0$-$8.0$ keV energy bands. We then intended to convolve our piecewise two-exponential model to the PSF profiles along each box for each energy range while fitting. However, the results we obtained were endowed with disproportionate errors due to the additional uncertainty brought by the spreading. Hence, we renounced taking the PSF directly into account in our fitting, and decided to present a ``worst case scenario'' to give an idea of the possible effects of the PSF on the results shown in Tables~\ref{tab:profilesFS}, \ref{tab:profilesRS} and \ref{tab:profilesother}.

To do so, we searched the box along which the PSF evolved the most, both spatially and with energy, which was the third box of our FS profiles, the furthest from the optical center. We then convolved the profiles of the three energy ranges PSF maps along this box with an infinitely thin filament, i.e. a Dirac function. The results were then fitted with our model, and the differences between the retrieved FWHMs for each energy range give an idea of the influence of the PSF in the worst scenario, with infinitely thin filaments in our worst box as regards to the PSF. 

The results are shown in Fig.~\ref{fig:PSF_test}, and it appears that the PSF can potentially have a significant influence. The FWHMs upstream are consistently larger upstream than downstream, and the FWHMs on both sides increase with energy. It is important to note that the PSF profiles all behave in a similar way: the spreading radius values increase both with energy and with the distance from the center (downstream to upstream for the forward shock, upstream to downstream for the reverse shock). Hence, the narrowing of the filaments with energy we observed might be underestimated because of the PSF, and our observations regarding the sharper decline downstream than upstream of the forward shock might be due to the PSF. The $m_E$ between the $0.4$-$1.7$ keV and $2.5$-$4.0$ keV profiles is $0.05 \pm 0.03$ and the $m_E$ between the $2.5$-$4.0$ keV and $5.0$-$8.0$ keV profiles is $0.22 \pm 0.08$, indicating that the PSF would tend to widen the filaments more at higher energies. 

\subsection{Interpretation}

The calculated $m_E$ values tend to indicate a stronger energy-dependence of the narrowing at high energies, which is likely underestimated by the PSF. According to \cite{Tran_2015}, this would be characteristic of a damping mechanism, where the rims widths are supposed to be energy-independent below a threshold energy, and decrease or increase once advection and/or diffusion controls the widths. Following the same paper, the action of a damping mechanism can be more confidently assessed by the observation of thin synchrotron filaments in radio. In the radio VLA observations of Cas~A, such filaments can be seen \citep[see for example][]{2014ApJ...785....7D}, which also suggests that damping effects are at stake in this SNR's synchrotron emission. A thorough joint study of both radio and X-ray filaments could allow for an estimation of the damping lengths.

\section{Conclusions}

From our study of the filament profiles in Cas~A, we can conclude the following~:

   \begin{enumerate}
      \item Our blind source separation method was able to produce clear, detailed and unpolluted maps of the synchrotron emission around three energy bands: $0.4$-$1.7$ keV, $2.5$-$4.0$ keV, and $5.0$-$8.0$ keV. These images clearly show filamentary structures all around the remnant, some associated to the forward shock, and some to the reverse shock. Some profiles can also be seen that cannot be clearly attributed to the forward or to the reverse shock, possibly because of line of sight effects.
      \item There is a noticeable narrowing of the synchrotron rims from lower to higher energies. Contrary to a previous study by \cite{Araya_2010}, we find that this effect is also visible between $2.5$-$4.0$ keV and $5.0$-$8.0$ keV, not only at lower energies. It is observed for filaments both at the forward and at the reverse shock, and for non-identified filaments. The evolution of the PSF with energy cannot account for this narrowing; on the contrary, the PSF would tend to make filaments appear wider with energy.
      \item There seems to be a stronger energy-dependence of this narrowing at higher energies, which would be a sign that there is a damping mechanism at stake. The observation of thin synchrotron profiles in the radio 6cm VLA observation of Cas~A is another indicator of this damping mechanism. However, it does not exclude the possible influence of the loss-limited effect as well.
      \item The filament profiles at the forward shock tend to be sharper downstream than upstream. At the reverse shock, the profiles mainly tend to be sharper upstream. However, this could be due to the radial degradation of the PSF away from the aim point. It could also be caused by projection effects that do not have the same impact on the forward and the reverse shock as they propagate in opposite directions.

   \end{enumerate}

\begin{acknowledgements}
The material is based upon work supported by NASA under award number 80GSFC21M0002. We thank D. Castro for the useful discussion about the PSF potential influence, and E. Costantini for the interesting discussion about potential dust scattering effects.
\end{acknowledgements}

%
%
\bibliographystyle{aa} 
\bibliography{synchrotron_casA}

\begin{thebibliography}{19}
\expandafter\ifx\csname natexlab\endcsname\relax\def\natexlab#1{#1}\fi

\bibitem[{Aharonian \& Atoyan(1999)}]{Aharonian09}
Aharonian, F.~A. \& Atoyan, A.~M. 1999, \aap

\bibitem[{Araya {et~al.}(2010)Araya, Lomiashvili, Chang, Lyutikov, \&
  Cui}]{Araya_2010}
Araya, M., Lomiashvili, D., Chang, C., Lyutikov, M., \& Cui, W. 2010, \apj,
  714, 396

\bibitem[{Bamba {et~al.}(2005)Bamba, Yamazaki, \& Hiraga}]{Bamba_2005}
Bamba, A., Yamazaki, R., \& Hiraga, J.~S. 2005, \apj, 632, 294

\bibitem[{Bobin {et~al.}(2020)Bobin, Hamzaoui, Picquenot, \& Acero}]{9215040}
Bobin, J., Hamzaoui, I.~E., Picquenot, A., \& Acero, F. 2020, IEEE Transactions
  on Image Processing, 29, 9429

\bibitem[{{Bobin} {et~al.}(2015){Bobin}, {Rapin}, {Larue}, \&
  {Starck}}]{bobin15}
{Bobin}, J., {Rapin}, J., {Larue}, A., \& {Starck}, J.-L. 2015, IEEE
  Transactions on Signal Processing, 63, 1199

\bibitem[{Cassam-Chenai {et~al.}(2007)Cassam-Chenai, Hughes, Ballet, \&
  Decourchelle}]{Cassam_Chenai_2007}
Cassam-Chenai, G., Hughes, J.~P., Ballet, J., \& Decourchelle, A. 2007, \apj,
  665, 315

\bibitem[{Corrales \& Paerels(2015)}]{10.1093/mnras/stv1704}
Corrales, L.~R. \& Paerels, F. 2015, Monthly Notices of the Royal Astronomical
  Society, 453, 1121

\bibitem[{Costantini \&
  Corrales(2022)}]{https://doi.org/10.48550/arxiv.2209.05261}
Costantini, E. \& Corrales, L. 2022, Interstellar absorption and dust
  scattering

\bibitem[{{DeLaney} {et~al.}(2014){DeLaney}, {Kassim}, {Rudnick}, \&
  {Perley}}]{2014ApJ...785....7D}
{DeLaney}, T., {Kassim}, N.~E., {Rudnick}, L., \& {Perley}, R.~A. 2014, \apj,
  785, 7

\bibitem[{{Helder} \& {Vink}(2008)}]{Helder08}
{Helder}, E.~A. \& {Vink}, J. 2008, \apj, 686, 1094

\bibitem[{{Hwang} {et~al.}(2004){Hwang}, {Laming}, {Badenes}, {Berendse},
  {Blondin}, {Cioffi}, {DeLaney}, {Dewey}, {Fesen}, {Flanagan}, {Fryer},
  {Ghavamian}, {Hughes}, {Morse}, {Plucinsky}, {Petre}, {Pohl}, {Rudnick},
  {Sankrit}, {Slane}, {Smith}, {Vink}, \& {Warren}}]{hwang04}
{Hwang}, U., {Laming}, J.~M., {Badenes}, C., {et~al.} 2004, \apjl, 615, L117

\bibitem[{Parizot {et~al.}(2006)Parizot, Marcowith, Ballet, \&
  Gallant}]{Parizot2006}
Parizot, E., Marcowith, A., Ballet, J., \& Gallant, Y.~A. 2006, \aap, 453, 387

\bibitem[{Picquenot {et~al.}(2019)Picquenot, Acero, Bobin, Maggi, Ballet, \&
  Pratt}]{picquenot:hal-02160434}
Picquenot, A., Acero, F., Bobin, J., {et~al.} 2019, A\&A, 627, A139

\bibitem[{Picquenot {et~al.}(2021)Picquenot, Acero, Holland-Ashford, Lopez, \&
  Bobin}]{Picquenot_2021}
Picquenot, A., Acero, F., Holland-Ashford, T., Lopez, L.~A., \& Bobin, J. 2021,
  \aap, 646, A82

\bibitem[{Pohl {et~al.}(2005)Pohl, Yan, \& Lazarian}]{Pohl_2005}
Pohl, M., Yan, H., \& Lazarian, A. 2005, \apj, 626, L101

\bibitem[{Ressler {et~al.}(2014)Ressler, Katsuda, Reynolds, Long, Petre,
  Williams, \& Winkler}]{Ressler_2014}
Ressler, S.~M., Katsuda, S., Reynolds, S.~P., {et~al.} 2014, \apj, 790, 85

\bibitem[{{Slane} {et~al.}(2014){Slane}, {Lee}, {Ellison}, {Patnaude},
  {Hughes}, {Eriksen}, {Castro}, \& {Nagataki}}]{2014ApJ...783...33S}
{Slane}, P., {Lee}, S.~H., {Ellison}, D.~C., {et~al.} 2014, \apj, 783, 33

\bibitem[{Tran {et~al.}(2015)Tran, Williams, Petre, Ressler, \&
  Reynolds}]{Tran_2015}
Tran, A., Williams, B.~J., Petre, R., Ressler, S.~M., \& Reynolds, S.~P. 2015,
  \apj, 812, 101

\bibitem[{Vink \& Laming(2003)}]{Vink_2003}
Vink, J. \& Laming, J.~M. 2003, \apj, 584, 758

\end{thebibliography}

\end{document}